\documentclass[review]{elsarticle}

\usepackage{hyperref}

\usepackage{epsfig}
\usepackage{amssymb}
\usepackage{mathptmx} 
\usepackage{amsmath}

\journal{Journal of \LaTeX\ Templates}

\newcommand{\bd}{\begin{displaymath}}
\newcommand{\ed}{\end{displaymath}}
\newcommand{\be}{\begin{equation}}
\newcommand{\ee}{\end{equation}}
\newcommand{\beq}{\begin{eqnarray}}
\newcommand{\eeq}{\end{eqnarray}}  
\newcommand{\beqs}{\begin{eqnarray*}}
\newcommand{\eeqs}{\end{eqnarray*}}  

\begin{document}

\begin{frontmatter}

\title{Towards rationalizing photoswitchable behavior of Cu$^{\mathrm{II}}_{2}$Mo$^{\mathrm{IV}}$ cyanido-bridged molecule}

\author[ad1]{Robert Pe\l ka\corref{corauth}}
\ead{robert.pelka@ifj.edu.pl}

\author[ad2,ad3]{Olaf Stefa\'{n}czyk}
\ead{olaf.stefanczyk@uj.edu.pl}

\author[ad4]{Anna Ma\l gorzata Majcher-Fitas}
\ead{anna.majcher@uj.edu.pl}

\author[ad5]{Corine Mathoni\`{e}re}
\ead{mathon@icmcb-bordeaux.cnrs.fr}

\author[ad3]{Barbara Sieklucka}
\ead{barbara.sieklucka@uj.edu.pl}

\cortext[corauth]{Corresponding author} 
\address[ad1]{The Henryk Niewodnicza\'{n}ski Institute of Physics, Polish Academy of Sciences, Radzikowskiego 152, 31-342 Krak\'{o}w, Poland}
\address[ad2]{Department of Chemistry, School of Science, The University of Tokyo, 7-3-1 Hongo, Bunkyo-ku, Tokyo 113-0033, Japan}
\address[ad3]{Faculty of Chemistry, Jagiellonian University, Gronostajowa 2, 30-387 Krak\'{o}w, Poland}
\address[ad4]{The Marian Smoluchowski Institute of Physics, Jagiellonian University, prof. St.\L ojasiewicza 11, 30-348 Krak\'{o}w, Poland}
\address[ad5]{ICMCB-CNRS, Universit\'{e} de Bordeaux, 87, Avenue du Docteur Schweitzer 33608 PESSAC cedex, France}

\begin{abstract}
[Cu$^{\mathrm{II}}$(enpnen)]$_{2}$[Mo$^{\mathrm{IV}}$(CN)$_{8}$]$\cdot$7H$_{2}$O (enpnen = N,N$^\prime$ - bis(2- aminoethyl)- 1,3 - propane- diamine) molecular cluster compound was subject to a series of irradiations with the light of 405 nm. On irradiation isothermal magnetization at 1.8 and 5 K in the field range 0 - 70 kOe as well as magnetic susceptibility in the temperature range of 2 - 300 K were subsequently detected. Both types of magnetic signals were next analyzed assuming that the irradiation triggers two independent processes: the metal to metal charge transfer (MMCT) leading to a state with the Arrhenius-type relaxation and the spin crossover (SC) transition ending in a state whose relaxation displays a threshold behavior. The first mechanism leads to an electron from the spinless Mo(IV) configuration being transferred to one of the Cu(II) ions transforming the trimer into the state Cu(II)-N-C-Mo(V)-C-N-Cu(I), with spin 1/2 on the Mo(V) ion and the spinless Cu(I) ion. The other mechanism gives rise to an excited paramagnetic Mo(IV)$^{*}$ linked to two paramagnetic Cu(II) centers with a possible superexchange interaction. The spin of the excited Mo(IV)$^{*}$ species is equal to 1 and associated to a disruption of the 5s-electronic pair. A reasonable result of simultaneous fitting the full series of susceptibility and magnetization data to the model taking into account both mechanisms corroborates their presence. 
\end{abstract}


\end{frontmatter}


\section{Introduction}

Spins and magnetization of molecular materials can be uniquely controlled by photons, which is the feature absent for conventional magnets \cite{Sato,Munoz,Bleuzen,Sieklucka,Ohkoshi,Funck}. The spin state of a molecular magnet can be changed either through photo-induced electron transfer (metal-to-metal charge-transfer MMCT process), where the redox states of two different metal centers are modulated \cite{Aguila,Zhang,Dei}, or through the light-induced excited spin state trapping (LIESST) effect (photo induced transformation from low spin LS to high spin HS state) \cite{Guionneau,Letard}. The studied cluster compound [Cu$^{\mathrm{II}}$(enpnen)]$_{2}$[Mo$^{\mathrm{IV}}$(CN)$_{8}$]$\cdot$7H$_{2}$O reveals both mechanisms, which was reported in detail in \cite{Stefanczyk1}. In the latter work a model was put forward which enabled the calculation of the temperature dependence of the photo-induced susceptibility and the field-dependence of the isothermal magnetization detected just upon irradiation. The present work aims basically at repeating this approach. However, two crucial modifications are introduced. Firstly, unlike in \cite{Stefanczyk1}, where only the susceptibility data were taken into account in the fitting procedure, the full set of magnetic data (both susceptibility and magnetization) are incorporated into the test function to be minimized. Secondly, here the magnetic properties of the compound are calculated assuming separate spectroscopic factors for the MMCT and SC excited states of the Mo ion. This feature was absent from the previous model, where a single averaged spectroscopic factor was used to describe the field-dependence of both excited states of the cluster compound. In addition, while this approach to rationalizing the photomagnetic properties of a molecular compound is relatively novel (it has been sketched for the first time in \cite{Korzeniak1}, its crucial details were hidden to a large extent in the Supporting Information file of \cite{Stefanczyk1} and might therefore be easily overlooked by a less diligent reader. We thus thought that a complete presentation in a dedicated article may come in right and useful.

\section{Experimental}

A preliminary step in the investigations of the magnetic properties of compound \textbf{1} was the measurement of the isothermal magnetization and susceptibility of a bulk sample. The mass of 10.291 mg of the powder sample of 1 was mounted in a plastic draw in the Quantum Design SQUID probe. The magnetization was detected at temperatures 1.8, 3, 5, 8 K in the field range of 0-70 kOe and at 100 K in the field range of 0-40 kOe. The susceptibility was measured using the dc field of 10 kOe in the temperature range 1.8-300 K \cite{Stefanczyk1}.

\begin{figure}[h!]
\centering
\includegraphics[width=\textwidth]{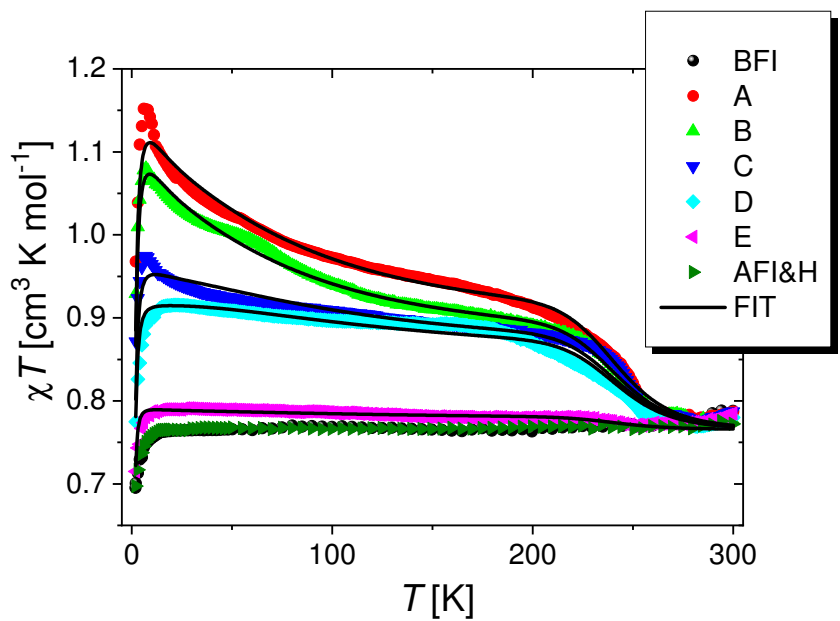}
\caption{Temperature dependence of $\chi T$ of \textbf{1} for all measurement series.}
\label{fig1}
\end{figure}

\begin{figure}[h!]
\centering
\includegraphics[width=\textwidth]{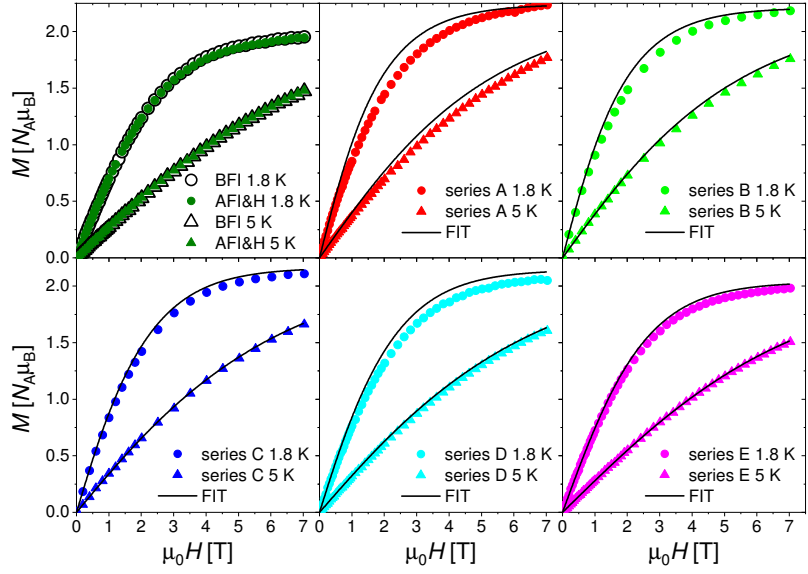}
\caption{Field dependence of the isothermal magnetization of \textbf{1} for all measurement series.}
\label{fig2}
\end{figure}

The main step of investigations consisted in irradiation of the sample with the visible light. The powder sample of \textbf{1} was scattered over a piece of scotch tape and mounted in the Quantum Design SQUID probe equipped with an optical fiber entry enabling the transmission of laser light. The first measurement of the sample was carried out prior to irradiation in two modes: for isothermal magnetization at 1.8 and 5 K in the field range of 0-70 kOe  (denoted by BFI in Fig. \ref{fig2}) and for susceptibility in the applied field of 10 kOe within temperature range of 1.8-300 K in the heating direction (denoted by BFI in Fig. \ref{fig1}). Next followed a series of subsequent runs in which the sample was first cooled down to 10 K (run A), 50 K (run B), 100 K (run C), 150 K (run D), and 200 K (run E) in the liquid helium bath and stabilized at these temperatures. Then the sample was irradiated with the purple light of 405 nm for as long as 33.38 h (run A), 24.50 h (run B), 24.56 h (run C), 24.43 h (run D), and 17.94 h (run E). During irradiation the $\chi T$ value was steadily increasing attaining the values of 1.133 emu K mol$^{-1}$ (run A), 1.003 emu K mol$^{-1}$ (run B), 0.905 emu K mol$^{-1}$ (run C), 0.892 emu K mol$^{-1}$ (run D), and 0.777 emu K mol$^{-1}$ (run E) just before switching off the light source. The relative increase in the signal between the initial state before irradiation and the final state after irradiation amounts to 50.9 \% (run A), 30.6 \% (run B), 17.7 \% (run C), 16.9 \% (run D), and 1.70 \% (run E). Just after stopping the irradiation the temperature of the sample was lowered down to 1.8 K and first the isothermal magnetization at 1.8 K and 5 K was measured in the field range of 0-70 kOe (increasing field mode) and immediately thereafter the field was fixed at the value of 10 kOe and the evolution of $\chi T$ was detected in the temperature range of 2-300 K. The temperature change during the susceptibility measurement mode was checked to be perfectly linear with a steady rate of 0.4 K min$^{-1}$. After each run the sample was cooled down to the initial temperature of the subsequent run. On finishing the final run with irradiation the sample was cooled down to liquid $^{4}$He, the magnetization was measured at 1.8 and 5 K, and finally the susceptibility was detected in the dc field of 10 kOe in the temperature range of 2-300 K (AFI\& H in Fig. \ref{fig1} and \ref{fig2}).

\section{Theory}

To get a more quantitative insight into the details of possible interactions in the unexcited state model "0" is assumed, where the intermolecular interaction between the trimer entities Cu$_{2}$Mo is treated within the molecular field theory. The corresponding Hamiltonian accounts only for the Zeeman coupling of the spins of the Cu(II) ions with the applied magnetic field
\be
\hat{\mathcal{H}}_0=g_\mathrm{Cu}\mu_\mathrm{B}\left(\hat{S}_{\mathrm{Cu}_1}+\hat{S}_{\mathrm{Cu}_2}\right)\cdot\vec{H}
\label{eq1}
\ee 
It is straightforward to derive the field- and temperature dependent partition function $Z_0$ related to that Hamiltonian:
\be
Z_0(H,T)=2\left[1+\cosh\left(\beta\mu_\mathrm{B}g_\mathrm{Cu}H\right)\right],
\label{eq2}
\ee
where $\beta=1/k_\mathrm{B}T$. The magnetization and susceptibility is obtained as appropriate derivatives of the partition function, i.e.
\beq
M_0(H,T)=\frac{N_\mathrm{A}}{\beta}\frac{\partial\ln Z_0(H,T)}{\partial H} &  & \chi_0(H,T)=\frac{N_\mathrm{A}}{\beta}\frac{\partial^2\ln Z_0(H,T)}{\partial H^2}.
\label{eq3}
\eeq 
It is easy to demonstrate that the magnetization function reduces to a sum of two Brillouin functions corresponding to a couple of the Cu(II) spins. The molar magnetic susceptibility was calculated according to the following approximate formula
\be
\chi_{0\mathrm{C}}=\frac{\chi_0}{1-\frac{z J^\prime\chi_0}{N_\mathrm{A}\mu_\mathrm{B}^2g_\mathrm{Cu}^2}},
\label{eq4}
\ee
where $\chi_{0\mathrm{C}}$ denotes the susceptibility corresponding to the interacting (coupled) clusters. The susceptibility calculated along the lines sketched above was fitted to the experimental points yielding ferromagnetic intermolecular coupling $z J^\prime=0.13(1)$\,cm$^{-1}$, and the Landé factor $g_\mathrm{Cu}=2.0168(8)$. The corresponding agreement quotient for susceptibility is  $R_{\chi T}=\sum\left[\left(\chi T\right)_\mathrm{obs}-\left(\chi T\right)_\mathrm{calc}\right]^2/\sum\left[\left(\chi T\right)_\mathrm{obs}\right]^2=4.1\cdot 10^{-5}$. Next, using the Brillouin function and relaxing the value of the copper Landé factor $g_\mathrm{Cu}$ the magnetization was calculated and fed into the analogous magnetization agreement quotient $R_M=\sum\left[M_\mathrm{obs}-M_\mathrm{calc}\right]^2/\sum\left[M_\mathrm{obs}\right]^2 $ incorporating all the available experimental points, which was optimized for a single common value of $g_\mathrm{Cu}$. The corresponding procedure was implemented in the \textit{Mathematica8.0} environment and yielded $g_\mathrm{Cu}=2.02(2)$ and $R_M=2.2\cdot 10^{-4}$. The sizeable value and the ferromagnetic character of the intermolecular coupling may point to the presence of intramolecular interaction between the Cu(II) ions mediated through the diamagnetic -N-C-Mo(IV)-C-N- linkage. The set of parameters $z J^\prime=0.13(1)$\,cm$^{-1}$ and  $g_\mathrm{Cu}=2.02(2)$ will serve in what follows as an input for the analysis of the susceptibility signal upon irradiation.

The irradiation of the sample in each run (from A to E) resulted in an apparent increase of the $\chi T$ values (see Fig. \ref{fig1}). At the lowest temperatures all the curves display consistently a downturn, which to a large extent may be due to the substantial dc field (10 kOe) present during the measurement, but the antiferromagnetic intermolecular coupling is not excluded. For runs A to C three regimes are apparent in the flow of the $\chi T$ curves on increasing the temperature. The first regime is characterized by a rather steep fall of the $\chi T$ values followed by a more gentle decrease (second regime) and finally an abrupt drop reaching the level detected before irradiation (third regime). Runs D and E lack the first regime, while maintaining the other two ones. The overall increase of the $\chi T$ signal upon irradiation may originate from two independent mechanisms. The first is the metal to metal charge transfer (MMCT), where an electron from the spinless Mo(IV) configuration is transferred to one of the Cu(II) ions transforming the trimer into the state Cu(II)-N-C-Mo(V)-C-N-Cu(I), with spin 1/2 on the Mo(V) ion and the spinless Cu(I) ion. The spin carriers in thus excited state may be exchange-coupled through the diamagnetic cyanido bridge. The other is a spin crossover (SC) process centered on the Mo(IV) ion due to irradiation in the LF region of the [Mo$^{\mathrm{IV}}$(CN)$_{8}$]$^{4-}$ spectra, giving rise to an excited paramagnetic Mo(IV)$^{*}$ linking two paramagnetic Cu(II) centers with a possible superexchange interaction. The spin of the excited Mo(IV)$^{*}$ species is equal to 1 and associated to a disruption of the 5s-electronic pair. It is important to note two facts. Firstly, the Cu(II)-N-C-Mo(V)-C-N-Cu(I) state as well as the Cu(II)-N-C-Mo(IV)$^{*}$-C-N-Cu(II) state are excited and thus metastable, and with the course of time they will relax back to the ground state of the trimer. Secondly, only a certain part of the trimer molecules in the sample shall undergo either an MMCT process or a SC process leaving the remaining molecules in the unperturbed state described by model "0".

\begin{figure}[h!]
\centering
\includegraphics[width=0.5\textwidth]{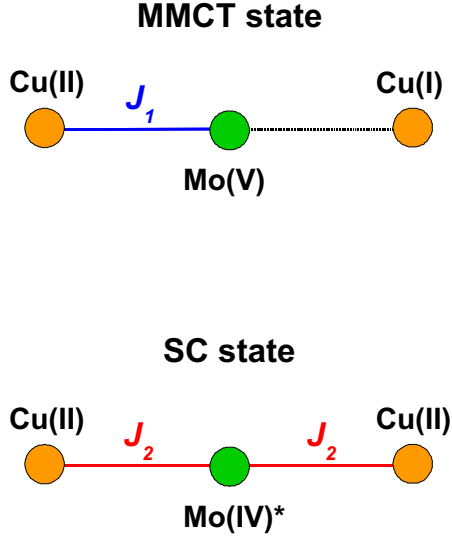}
\caption{Schematics of models I and II of MMCT (top) and SC (bottom) excited states.}
\label{fig3}
\end{figure}

Pertinent to the new photoinduced configurations of the trimer molecule we assume model I and II depicted in Fig. \ref{fig3}, where $J_1$ and $J_2$ denote superexchange constants of the coupling between, respectively, the spin 1/2 state of the Mo(V) ion and the spin 1/2 state of the remaining Cu(II) ion and the spin 1 state of the Mo(IV)$^{*}$ ion and the spin 1/2 states of each of the Cu(II) ions. One comment is necessary here. Although the unit cell of \textbf{1} contains two symmetry inequivalent trimer entities, the bond lengths and angles in the cyanido bridges mediating the superexchange coupling differ only little between both entities and between either side of each entity. Therefore, we resort to a natural simplification consisting in assuming that all the trimers in compound \textbf{1} are structurally identical. Hence our models involve a single exchange coupling constant each. This assumption reduces considerably the number of free parameters to be determined. 
The Hamiltonian corresponding to model I acquires the following form
\be
\hat{\mathcal{H}}_1=-J_1\hat{S}_{\mathrm{Mo}_1}\cdot\hat{S}_\mathrm{Cu}+\mu_\mathrm{B}\left(g_{\mathrm{Mo}_1}\hat{S}_{\mathrm{Mo}_1}+g_\mathrm{Cu}\hat{S}_\mathrm{Cu}\right)\cdot\vec{H},
\label{eq5}
\ee
where different Land\'{e} factors for the Mo(V) and Cu(II) ions are assumed and $J_1$ is the superexchange coupling constant between spin 1/2 of the Mo(V) ion and spin 1/2 of the unperturbed Cu(II) ion. It is straightforward to derive the field- and temperature dependent partition function $Z_1$ related to Hamiltonian  $\hat{\mathcal{H}}_1$:
\beq
Z_1(\beta,H)&=&2e^{\frac{1}{4}\beta J_1}\cosh\left[\frac{1}{2}\beta\mu_\mathrm{B}(g_\mathrm{Cu}+g_{\mathrm{Mo}_1})H\right] \nonumber\\
   &+&2e^{-\frac{1}{4}\beta J_1}\cosh\left[\frac{1}{2}\beta\sqrt{\mu_\mathrm{B}^2(g_\mathrm{Cu}-g_{\mathrm{Mo}_1})^2H^2+J_1^2}\right]\label{eq6}
\eeq
The Hamiltonian corresponding to model II reads
\be
\hat{\mathcal{H}}_2=-J_2\hat{S}_{\mathrm{Mo}_2}\cdot\left(\hat{S}_{\mathrm{Cu}_1}+\hat{S}_{\mathrm{Cu}_2}\right)+\mu_\mathrm{B}\left[g_\mathrm{Cu}\left(\hat{S}_{\mathrm{Cu}_1}+\hat{S}_{\mathrm{Cu}_2}\right)+g_{\mathrm{Mo}_2}\hat{S}_{\mathrm{Mo}_2}\right]\cdot\vec{H}
\label{eq7}
\ee
where similarly different Land\'{e} factors for the Mo(IV)$^{*}$ and Cu(II) ions were assumed and $J_2$ is the superexchange coupling constant between spin 1 of the Mo(IV)$^{*}$ ion and spins 1/2 of both the Cu(II) ions. In this case the corresponding partition function $Z_2$ cannot be given in a closed form and must be calculated numerically through the diagonalization of the Hamiltonian in Eq. {\ref{eq7}} for each required value of the couple $(H,T)$. The molar magnetization and susceptibility in both models is obtained as appropriate derivatives of the partition functions. We deliberately decide not to take into account the intermolecular interactions in both models in order to simplify the overall description and reduce the number of free parameters. Let us finally note that the two Land\'{e} factors $g_{\mathrm{Mo}_1}$ and $g_{\mathrm{Mo}_2}$ correspond to two distinct excited states of the Mo ion, i.e. Mo$_1$=Mo(V) and Mo$_2$=Mo(IV)$^{*}$, and as such they are in general not equal.

As mentioned above only a fraction $f_1$ of the sample undergoes the MMCT process giving rise to the enhanced susceptibility signal. The excited MMCT state is a metastable one, so the fraction $f_1$ diminishes in the course of time. Let us assume that the number of the MMCT excited trimer units and concomitantly with it the fraction $f_1$ decreases with time according to the exponential (Arrhenius) law given by
\be
f_1=f_{10}\exp\left(-\frac{t}{\tau}\right),
\label{eq8}
\ee
where $f_{10}$ is the initial fraction at moment $t=0$ and $\tau$ is the mean lifetime of the metastable state revealing a temperature dependence associated with the presence of the energy barrier $\Delta_1$ preventing the state from relaxing back to the ground state described by an exponential rule
\be
\tau=\tau_0\exp\left(\frac{\Delta_1}{k_\mathrm{B}T}\right),
\label{eq9}
\ee
where $\tau_0$ defines a high-temperature limit of the lifetime. In the measurement performed on laser light irradiation the temperature of the sample increases linearly from $T_\mathrm{min}=2$\,K to $T_\mathrm{max}=300$\,K within the period of $\Delta t_\mathrm{m}=$12 h 25 min (the rate of increase is 0.4 K min$^{-1}$), so one can write a simple relation between the current time $t$ (where $t=0$ coincides with the start of the measurement) and the current temperature of the sample
\be
t=\frac{T-T_\mathrm{min}}{T_\mathrm{max}-T_\mathrm{min}}\Delta t_\mathrm{m}.
\label{eq10}
\ee 
Thus the fraction of the MMCT excited trimer units in the sample $f_1$ is the following function of temperature
\be
f_1=f_{10}\exp\left[-e^{-\frac{\Delta_1}{k_\mathrm{B}T}}\frac{T-T_\mathrm{min}}{T_\mathrm{max}-T_\mathrm{min}}\rho\right],
\label{eq11}
\ee
where $\rho=\Delta t_\mathrm{m}/\tau_0$.

Similarly, only a fraction $f_2$ of the sample undergoes the SC process. In this case we expect that the deexcitation of the SC state and its return to the ground state configuration is a threshold process, where up to a certain temperature $T_0$ the population of the SC state remains roughly on the same level. It is only above the threshold temperature $T_0$ that the deexcitation process starts to be effective and the fraction $f_2$ starts to diminish noticeably. For the precise shape of temperature dependence of this process we draw on the formula describing the reverse SC transition \cite{Kahn}, i.e.
\be
f_2=\frac{f_{20}}{1+\exp\left[\Delta_2\left(\frac{1}{T_0}-\frac{1}{T}\right)\right]},
\label{eq12}
\ee
where $f_{20}$ denotes the initial (at the start of the measurement) fraction of the SC excited trimer units and $\Delta_2$ is a parameter accounting for the energy distance between the excited and ground state. Finally, the fraction of the ground state $f_0$ obeys at each temperature the following relation  $f_0=1-f_1-f_2$. Now, the susceptibility signal detected just upon irradiation denoted with the subscript "UI" can be described by an appropriate combination of signals inferred from model "0" (not excited trimer units) and model I (MMCT excited trimer units) and model II (SC excited trimer units):
\beq
\chi_\mathrm{UI}(H,T)&=&\left[1-f_1\left(f_{10},\Delta_1,\rho,T\right)-f_2\left(f_{20},\Delta_2,T_0,T\right)\right]\chi_{0\mathrm{C}}\left(z J^\prime,g_\mathrm{Cu},H,T\right)\nonumber \\
 &+& f_1\left(f_{10},\Delta_1,\rho,T\right)\chi_1\left(J_1,g_\mathrm{Cu},g_{\mathrm{Mo}_1},H,T\right)\nonumber \\
 &+& f_2\left(f_{20},\Delta_2,T_0,T\right)\chi_2\left(J_2,g_\mathrm{Cu},g_{\mathrm{Mo}_2},H,T\right)
\label{eq13}
\eeq
where all the parameters and variables are given explicitly. The total magnetization upon irradiation was measured in each instance prior to the susceptibility run, hence one should calculate it as an analogous combination of signals using the initial values of the fractions, i.e.
\beq
M_\mathrm{UI}(H,T)&=&\left[1-f_{10}-f_{20}\right]M_0\left(z J^\prime,g_\mathrm{Cu},H,T\right)\nonumber \\
 &+& f_{10}M_1\left(J_1,g_\mathrm{Cu},g_{\mathrm{Mo}_1},H,T\right)\nonumber \\
 &+& f_{20}M_2\left(J_2,g_\mathrm{Cu},g_{\mathrm{Mo}_2},H,T\right).
\label{eq14}
\eeq

\section{Results and Discussion}

The main step of the analysis is the modeling of the susceptibility and magnetization signals observed upon irradiation in the five subsequent runs from A to E. To this end we defined a test function $Q$ being the sum of two agreement quotients $Q_{\chi T}$ and $Q_M$ ($Q=Q_{\chi T}+Q_M$) taking into account all the measurement runs, where
\be
Q_{\chi T}=\frac{\sum_{\omega\in\Omega}\sum_{i_\omega}\left[\left(\chi T\right)_{i_\omega}-\chi_\mathrm{UI}\left(H,T_{i_\omega}\right)T_{i_\omega}\right]^2}{\sum_{\omega\in\Omega}\sum_{i_\omega}\left[\left(\chi T\right)_{i_\omega}\right]^2}
\label{eq15}
\ee
and
\be
Q_M=\sum_{i=1}^{2}\frac{\sum_{\omega\in\Omega}\sum_{j_\omega}\left[M_{i,j_\omega}-M_\mathrm{UI}\left(H_{j_\omega},T_i\right)\right]^2}{\sum_{\omega\in\Omega}\sum_{j_\omega}\left[M_{i,j_\omega}\right]^2}
\label{eq16}
\ee
where $\Omega=\left\{\mathrm{A,B,C,D,E}\right\}$ is a set of the measurement runs, indices $i_\omega$ and $j_\omega$ enumerate the consecutive measurement points in a given run, and index $i=1,2$ corresponds to the two temperatures ($T_1=1.8$\,K and $T_2=5$\,K) at which the magnetization was measured in each run. Let us note that each run is characterized by an independent set of the initial fractions \{$f_{\omega 00}$, $f_{\omega 10}$, $f_{\omega 20}$\} of which only two are independent variables ($f_{\omega 00}+f_{\omega 10}+f_{\omega 20}=1$). Parameters $z J^\prime$  and $g_\mathrm{Cu}$ were fixed at the values found in the analysis of the bulk data. This leaves us with a set of 18 independent parameters: $\Delta_1$, $\rho$, $\Delta_2$, $T_0$, $J_1$, $J_2$, $g_{\mathrm{Mo}_1}$, $g_{\mathrm{Mo}_2}$ $f_{\omega 00}$, $f_{\omega 10}$ ($\omega\in\Omega$). Let us note here that the data set to be analyzed is quite large with 94 experimental points falling on one parameter. Using the implementation of the above definition in the \textit{Mathematica8.0} environment function $Q$ was minimized by the principal axis method. Unlike in \cite{Stefanczyk1}, where only after fixing $f_{\mathrm{D}10}=f_{\mathrm{E}10}=0$ the optimizing procedure converged returning an acceptable solution, it did converge in its full yielding reasonable parameter values. The minimum thus obtained amounts to $Q=8.6\cdot 10^{-5}$ ($Q_{\chi T}=8.1\cdot 10^{-5}$ and $Q_M=5.5\cdot 10^{-6}$) and corresponds to the set of parameters listed in Table \ref{tab1}. This result should be compared to that reported in \cite{Stefanczyk1}, where $Q_{\chi T}=6.4\cdot 10^{-5}$ is in fact slightly smaller than the present value, however $Q_M=5.3\cdot 10^{-3}$ differs by three orders of magnitude from the present result.
\begin{table}[h!]
\begin{center}
\begin{tabular}{|l|cccccc|}
\hline\hline
parameter & $J_1$ [cm$^{-1}$] & $J_2$ [cm$^{-1}$] & $\Delta_1$ [K] & $\rho$ & $\Delta_2$ [K] & $T_0$ [K] \\
value & 17 & 104 & 81 & 7 & 4306 & 243 \\
relative error [\%] & 55 & 0.19 & 90 & 76 & 59 & 3.4 \\
\hline\hline
parameter & $g_{\mathrm{Mo}_1}$ & $g_{\mathrm{Mo}_2}$ & $f_{\mathrm{A}00}$ & $f_{\mathrm{A}10}$ & $f_{\mathrm{B}00}$ & $f_{\mathrm{B}10}$ \\
value & 2.22 & 2.25 & 0.471 & 0.451 & 0.504 & 0.431 \\
relative error [\%] & 4.1 & 0.18 & 1.4 & 1.5 & 1.3 & 1.5 \\
\hline\hline
parameter & $f_{\mathrm{C}00}$ & $f_{\mathrm{C}10}$ & $f_{\mathrm{D}00}$ & $f_{\mathrm{D}10}$ & $f_{\mathrm{E}00}$ & $f_{\mathrm{E}10}$ \\
value & 0.841 & 0.100 & 0.937 & 0.008 & 0.992 & 0.0005 \\
relative error [\%] & 0.77 & 6.7 & 0.69 & 80 & 0.65 & $>100$ \\
\hline\hline 
\end{tabular}
\end{center}
\caption{Parameter values corresponding to the minimum of $Q$\label{tab1}}
\end{table}
The relative errors listed in Table \ref{tab1} reveal that the set of the best fit parameters breaks up into two groups: the group of six parameters ($J_1$, $\Delta_1$, $\rho$, $\Delta_2$, $f_{\mathrm{D}10}$, $f_{\mathrm{E}10}$) to which function $Q$ is only weakly sensitive (considerable relative errors) and the remaining twelve parameters ($T_0$, $g_{\mathrm{Mo}_1}$, $g_{\mathrm{Mo}_2}$ and the remaining fractions $f_{\omega 00}$ and $f_{\omega 10}$) with fairly strong influence on $Q$ (small relative errors). Let us at the same time stress that the relative errors within the first group do not exceed 100 \% but for the fraction $f_{\mathrm{E}10}$ which has the smallest absolute value. The exchange couplings in both the MMCT excited state and the SC excited state turn out to be of ferromagnetic character, which is consistent with the general increase of the susceptibility signal detected upon irradiation in all the measurement runs. Parameter $\rho$ implies that $\tau_0\approx 1.77$\,h, which is a reasonable result comparing well with the time scale of the photomagnetic experiment. Moreover, the height of the energy barrier for the relaxation of the MMCT excited state $\Delta_1\approx 80$\,K is also plausible. Parameter $\Delta_2\approx 4300$\,K constitutes a fraction (12\%) of the energy of the light quantum ($\approx 35430$\,K for the purple light with $\lambda = 405$\,nm) used in the irradiation of the sample, which should be expected. The values of the Land\'{e} factors $g_{\mathrm{Mo}_1}=2.22$ and $g_{\mathrm{Mo}_2}=2.25$ are comparable and consistent with the upper bound of the interval [2.13,2.24] reported in \cite{Stefanczyk1}. Finally, the fractions $f_{\omega 00}$ and $f_{\omega 10}$ ($\omega\in\Omega$) are all positive and attain consistently values below 1. 

\begin{figure}[h!]
\centering
\includegraphics[width=\textwidth]{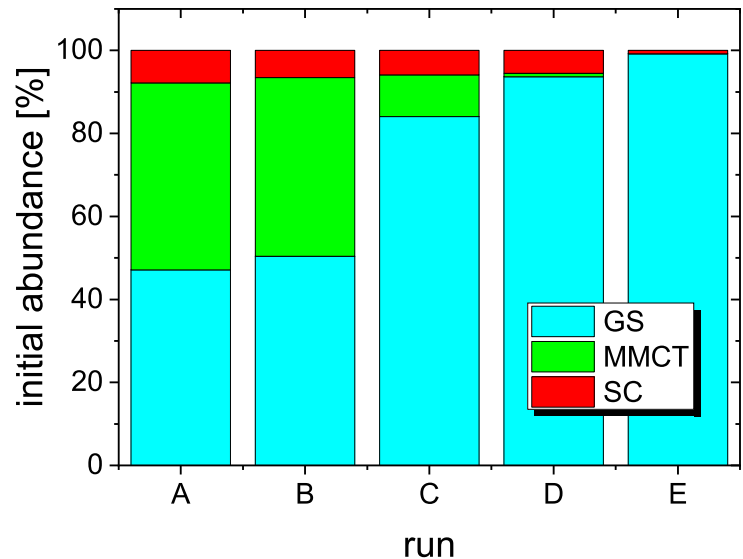}
\caption{Initial abundances of the ground state (GS) and the MMCT and SC excited states for all the measurement runs A-E.}
\label{fig4}
\end{figure}

The curves corresponding to the parameter values in Table \ref{tab1} are depicted with solid lines in Fig. \ref{fig1} (susceptibility) and Fig. \ref{fig2} (magnetization). The agreement of the calculated values (solid lines) with the experimental ones (symbols) is satisfactory although not perfect. The largest discrepancies in susceptibility (see Fig. \ref{fig1})can be observed in the low temperature limit, which may be due to the fact that for the sake of simplicity we neglected the intermolecular interactions between the clusters in their excited states. The introduction of two further parameters would certainly have an adverse effect on the convergence of the optimizing procedure. The agreement of the calculated values (solid lines) with the experimental ones (symbols) of the molar magnetization  (see Fig. \ref{fig2}) is apparently better at 5 K than at 1.8 K. At 1.8 K the calculated values overestimate the experimental signal especially for the intermediate field values. This again may be due to the lack of accounting for the intermolecular interactions of the timer units in their excited states, which most probably are antiferromagnetic in character. Another source of discrepancy may be the fact that the magnetization measurement preceded the susceptibility measurement in each run while the values of the initial fractions $f_{\omega 00}$ and $f_{\omega 10}$ ($\omega\in\Omega$) correspond precisely to the starting moment of the susceptibility measurement. However, this should rather result in a slight underestimation of the fractions and therefore lead to a further enhancement of the calculated values. For the sake of transparency of presentation the initial abundances (fractions) of the ground state and the MMCT and SC excited states are shown in Figure \ref{fig4} in a form of a stack column chart. It can be seen that the level of population of both the MMCT and SC states diminishes with increasing temperature of irradiation (along the series of the runs). The SC state has a nonvanishing contribution in all the runs, while the contribution of the MMCT state is negligibly small for the two final runs. Apparently, the time that elapses between the moment of irradiation finish and the moment of susceptibility measurement start, which gradually extends from run to run, is for the last two runs sufficiently long for the MMCT excited trimer units to relax back to the ground state.  

\section{Conclusions}

Let us note that the approach presented here represents our second attempt to quantitatively describe a magnetic photoinduced effect. A similar approach has been taken for the first time in \cite{Korzeniak1}, where two cluster compounds based on Cu(II) and Mo(IV) revealed a slight enhancement of their magnetic properties due to irradiation. Although the modeling of the photoinduced signals involved as many as eighteen parameters, the best fit parameter values seem physically acceptable. Moreover, the fact that both the susceptibility and the isothermal magnetization were reproduced to a satisfactory extent with these parameters bears out the correctness and reliability of the whole procedure. The initial fractions of excited trimer units diminish from run to run, which can be understood by remembering that the excited states are metastable. The successful analysis corroborates the assumption that the irradiation of the present sample triggers two independent processes: the metal to metal charge transfer (MMCT) leading to a state with the Arrhenius-type relaxation and the spin crossover (SC) transition ending in a state whose relaxation displays a threshold behavior. The extended procedure gave a unique insight into the values of the spectroscopic factors of both excited states of the Mo ion. They turned out to be comparable and exceed the free electron value, which indicates crucial anisotropy of both photo-generated states. 


\end{document}